\documentclass{aa}  

\usepackage{dcolumn}
\newcolumntype{d}[1]{D{.}{.}{#1}}  
\usepackage{graphicx}
\usepackage{xcolor,colortbl}
\usepackage{txfonts}
\usepackage{natbib}
\usepackage{stfloats}
\usepackage[colorlinks=true,urlcolor=blue,citecolor=blue,linkcolor=blue,breaklinks=true]{hyperref}
\begin{document}

   \title{Near-continuous tracking of solar active region NOAA\,13664 over three solar rotations}

   \author{I. Kontogiannis
          \inst{1,2}    
          \and
          Y. Zhu
          \inst{1}
          \and
          K. Barczynski
          \inst{3}
          \and
          M. Z. Stiefel
          \inst{1,4}
          \and
          H. Collier
          \inst{1,4}
          \and
          J. McKevitt
          \inst{5,6}
          \and
          J.S. Castellanos Dur\'{a}n
          \inst{7}
         \and
          S. Berdyugina
          \inst{2,8}
          \and
          L.K. Harra
          \inst{1,3}
          }
    \institute{ETH Z\"{u}rich, Institute for Particle Physics and Astrophysics, Wolfgang-Pauli-Strasse 27, 8093, Z\"{u}rich, Switzerland
           \and
           Istituto Ricerche Solari Aldo e Cele Dacc\`{o} (IRSOL), Faculty of Informatics, Universit\`{a} della Svizzera Italiana, CH-6605 Locarno, Switzerland
           \and
           PMOD/WRC, Dorfstrasse 33, 7260, Davos Dorf, Switzerland
           \and
           University of Applied Sciences and Arts Northwestern Switzerland, Bahnhofstrasse 6, 5210, Windisch, Switzerland
           \and
           University College London, Mullard Space Science Laboratory Holmbury St Mary, Dorking Surrey, RH5 6NT, UK
           \and
           University of Vienna, Institute of Astrophysics T\"{u}rkenschanzstrasse 17 Vienna A-1180, Austria
            \and
           Max Planck Institute for Solar System Research, Justus-von-Liebig-Weg 3, D-37077 G\"{o}ttingen, Germany
           \and
           Euler Institute, Faculty of Informatics, Universit\`{a} Svizzera Italiana, Lugano, Switzerland
            \\
              \email{ikontogianni@phys.ethz.ch}
        }
        
   \date{\today}

 
  \abstract
   {Magnetic flux emergence and decay in the Sun are processes that can span from days to months. However, their tracking is typically limited to about half a solar rotation when relying on single-vantage-point observations, providing only a partial view of the phenomenon.}
   {This study aims to monitor the magnetic and coronal evolution and characterise the non-potentiality of solar active region NOAA\,13664, one of the most complex and eruptive regions of the past two decades, over more than three full solar rotations, by combining observations from both the Earth-facing and far side of the Sun.}
   {We used photospheric magnetograms and Extreme Ultraviolet (EUV) filtergrams from Solar Orbiter and the Solar Dynamics Observatory, taken continuously over a 94-day period, along with 969 flare detections from combining the Geostationary Operational Environmental Satellite and the Spectrometer/Telescope for Imaging X-rays instrument on board the Solar Orbiter. All images were deprojected into a common coordinate system and merged into a unified dataset. We tracked the evolution of magnetic flux and EUV emission and computed magnetic field parameters from the line-of-sight magnetograms to quantify the region’s non-potentiality. The latter comprise the first continuous time series of their kind.}
   {We successfully identified the region’s initial emergence and followed its evolution through to its decay. The region developed through successive flux emergence episodes over a period of twenty days, reached its peak complexity one month after the first emergence, and gradually decayed over the subsequent two months. Unlike many complex regions, it consistently maintained high levels of non-potentiality for most of its lifetime, sustaining equally strong flaring activity. The derived time series of non-potentiality parameters far exceeded the typical 14-day window imposed by solar rotation and were remarkably consistent, exhibiting strong correlation with the flaring activity of the region. }
   {Multi-vantage-point observations offer valuable insights into the dynamics of flux emergence and decay, beyond the two-week limit imposed by solar rotation on observations along the Sun-Earth line. The corresponding combined datasets can significantly improve our understanding of how magnetic flux emerges, evolves, and drives solar activity.}


\titlerunning{NOAA\,13664 from emergence to decay}
\authorrunning{Kontogiannis et al.}
\maketitle
%

\section{Introduction}

Active regions on the Sun are areas associated with intense and complex magnetic fields. They are the primary sources of solar flares and coronal mass ejections (CMEs), which episodically release vast amounts of electromagnetic energy, solar plasma, and energetic particles into interplanetary space \citep{Fletcher11,webb12,GEORGOULIS2024}. The formation and evolution of these magnetic field concentrations drive dynamic processes throughout all layers of the solar atmosphere, often over timescales of weeks or even months \citep{2014SSRv..186..227S,2020AdSpR..65.1641I}. Consequently, understanding the formation and evolution of active regions is crucial for advancing our knowledge of solar magnetism and its influence on both localized and large-scale solar atmospheric conditions, as well as on the solar wind.

It is well established that active regions form through the interplay between the solar dynamo and magnetic buoyancy. As magnetic fields intensify near the base of the convection zone, they become buoyant and rise, undergoing deformation from turbulent convective motions \citep{cheung2008,cheung2010,2017ApJ...846..149C}. Upon reaching the photosphere, these fields organize into strong and extended concentrations, manifesting as groups of pores and sunspots. These are interconnected by magnetic loops extending into the upper atmosphere, with observable signatures appearing gradually across all atmospheric heights and temperatures. After the magnetic flux at the photosphere reaches its peak, a prolonged decay phase begins, gradually deforming the region and dispersing its magnetic flux \citep{2019A&A...625A..53S}. The total lifetime of active regions depends on their maximum size and surrounding magnetic environment, ranging from several days for smaller regions to months for larger ones. Typically, the emergence phase is much faster than the decay phase, lasting up to one-third of the region’s total lifetime \citep{2015LRSP...12....1V}.

While most active regions are relatively simple bipolar configurations, a subset evolve into extensive and highly complex magnetic configurations \citep{jaeggli_norton_2016}. Their magnetic field deviates significantly from a potential (current-free) configuration \citep{2016ApJ...820..103S}, thus these regions contain significant amounts of free magnetic energy and helicity and are the primary sources of the most intense solar eruptions \citep{2000ApJ...540..583S}. Despite progress in understanding flux emergence, the precise origin of complex active regions, particularly those containing $\delta$-spots, which are opposite polarity sunspots sharing a common penumbra \citep{1965AN....288..177K}, remains uncertain. These regions are known to exhibit systematically higher flux emergence rates \citep{norton22}, reach significantly greater peak magnetic flux and electric current densities, and take longer to evolve to their maximum extent \citep{2024ApJ...970..162K}. They may originate from the emergence of highly twisted, kink-unstable flux tubes, or result from interactions between distinct emerging flux systems, which may or may not be part of a single sub-photospheric structure \citep[see e.g.,][and references therein]{levens23,2019ApJ...871...67C}. Repeated emergence events and subsequent interactions between flux systems are common during the evolution of such regions \citep{toriumi17,kontogiannis2024}.

Moreover, statistically significant spatial persistence has been observed in the emergence of large and complex active regions, especially around solar maxima. These regions tend to form in nests or extended activity complexes that can persist for months, shaping the solar magnetic environment over multiple rotations \citep{1990SoPh..129..221B,2008ApJ...686.1432G}. However, the sub-photospheric mechanisms responsible for forming such nests and complex regions remain inaccessible to direct observation.

Advances in instrumentation, both ground-based and space-borne, have enabled regular monitoring of the solar photospheric magnetic field over extended periods, aiming to understand the evolution of active regions and activity complexes, their eruptive activity, and their contribution to coronal heating and solar magnetism. \citet{1993SoPh..147..269J} presented the long-term magnetic evolution and activity output of AR 5395, from February to April 1989, emphasizing on the region that produced the geomagnetic storm of March 1989. \citet{fisher1998} used magnetograms gathered over more than three years to investigate the scaling relationship between the coronal X-ray luminosity and the magnetic flux. These scalings were also examined during the evolution of an active region over five solar rotations \citep{2003ApJ...586..579V}.

Since the launch of the Solar Dynamics Observatory \citep[SDO;][]{sdo}, we have benefited from continuous, high-quality monitoring of the Sun’s magnetic field and corona across more than one full solar cycle \citep{2022SoPh..297...59K}. \citet{2017PASJ...69...47H} followed the changes in the magnetic topology of an active region over three solar rotations and demonstrated how the sites of upflowing plasma expand during the decay. \citet{2017IAUS..327...60C} presented the evolution of the magnetic flux and current helicity along with the CME output of a region over five solar rotations, while \citet{2020AdSpR..65.1641I} focused on specific activity periods and investigated the CME output in different evolutionary stages of the region. \citet{2017ApJ...840..100M} monitored the magnetic helicity evolution of active region NOAA\,12192 and its derivatives over three months, as well as the effect of the activity complex on the polar magnetic field. Following the evolution of activity complexes over several Carrington rotations \citet{2020ApJ...904...62W} pinpointed the impact of the active regions of the complexes to the polar magnetic field. \citet{2021RAA....21..312M} studied the magnetic properties, EUV emission, and flare productivity of a series of recurrent active regions over five months, highlighting how one of these regions, namely NOAA\,12673, stood out and produced the geomagnetic storm of September 2017. Notwithstanding their significance, our magnetic field observations have been largely limited to the Sun-Earth line, allowing continuous monitoring of any given region for only about 14 days. As a result, provided that larger active regions have lifetimes that exceed two weeks, key stages in their emergence, development, or decay are often missed, leading to an incomplete picture of their evolution.

The Solar Orbiter mission \citep{solo,zouganelis2020}, with its varying vantage point, offers a unique opportunity to overcome these limitations. Under favorable orbital conditions, it enables quasi-continuous monitoring of active regions, helping to bridge observational gaps. The Polarimetric and Helioseismic Imager \citep[PHI;][]{solo_phi} onboard Solar Orbiter provides regular magnetic field measurements in both full-disk and high-resolution modes. \citet{2024A&A...685A..28M} demonstrated that measurements from PHI’s full-disk telescope (FDT) align well with those from the Helioseismic and Magnetic Imager \citep[HMI;][]{hmischerrer,hmischou}. Additionally, far-side PHI magnetograms are consistent with helioseismic inferences \citep{2023A&A...674A.183Y}. Under specific geometric configurations, simultaneous observations from PHI and HMI can enhance our ability to reconstruct the vector magnetic field at the photosphere \citep{2023A&A...677A..25V}. These multi-vantage point observations extend the temporal coverage of photospheric magnetic field monitoring \citep{2024A&A...682A.108L}, enabling more comprehensive analyses of active region and activity complex evolution \citep{Finley2025}.

\begin{figure*}
   \centering
   \includegraphics[width=18cm]{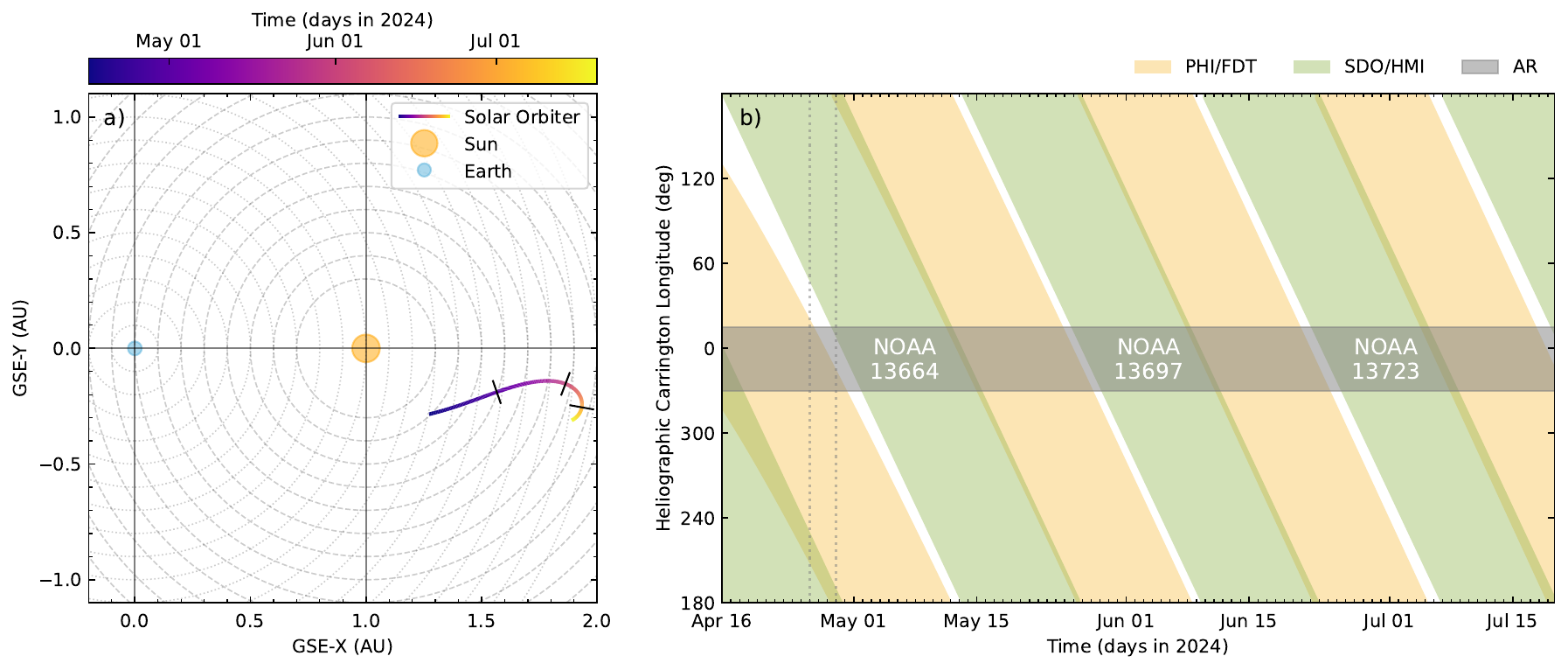}
      \caption{Left: The orbit of Solar Orbiter with respect to the Earth in Geocentric Solar Ecliptic (GSE) coordinates from 16 April to 18 July 2024. Right: Temporal coverage distribution of the region between the two instruments, The region was within the SO/PHI-FDT (HMI) FOV during the yellow (green) intervals. During each revolution, the region was assigned a different NOAA number, as indicated in the figure. Indicatively, we mark the observing gap between 26 and 29 April with the two vertical dotted lines. Similar gaps are also found later. }
        \label{fig:orbits}
   \end{figure*}

This study focuses on NOAA Active Region 13664, one of the most geoeffective regions of Solar Cycle 25, and its further evolutionary stages, designated NOAA\,13697 and 12723. The region appeared on the Earth-facing side of the Sun in late April 2024 and quickly evolved into one of the most magnetically complex active regions ever observed. Between 8 and 11 May 2024, it produced multiple X-class flares and associated CMEs, culminating in one of the most intense geomagnetic storms on record. Its complexity (non-potentiality) was driven by interactions among at least three intrinsically complex subregions. The region exhibited highly sheared magnetic fields with the highest peak values and injection rates of electric currents of the past two solar cycles. These were fueled by persistent motions and interactions among flux systems \citep{kontogiannis2024,2024ApJ...973L..31R}. Its exceptional non-potentiality was confirmed by multiple quantitative indicators \citep{Jaswal2025}, all pointing to the presence of numerous strong magnetic polarity inversion lines (MPILs), key sites of magnetic interaction that underpinned its extreme flare productivity \citep{Jarolim2024,mactaggart2025,wangrui2024}. Taking advantage of the favorable positioning of the Solar Orbiter between April and July 2024 (Fig.~\ref{fig:orbits}), we combined Solar Orbiter and SDO data to monitor the region near-continuously for 94 days, from its first emergence up to its decay.

\section{Data and Analysis}
\label{Section:analysis}
\subsection{Two vantage-point observations}

The Full Disk Telescope (FDT) of the PHI on board Solar Orbiter provides full-disk magnetograms of the Sun, utilizing the Fe~{\sc i} 6173\,\AA\ spectral line. Observations are obtained by sequentially scanning six wavelengths across the line profile in four polarization states. Magnetic field products are derived through a Milne-Eddington inversion \citep{2007A&A...462.1137O,2011SoPh..273..267B}. For this study, we use data from the third FDT data release\footnote{\url{https://www.mps.mpg.de/solar-physics/solar-orbiter-phi/data-releases}}, which includes only the line-of-sight (LOS) component of the magnetic field, $B_{LOS}$. This is sufficient for our purposes, as the $B_{LOS}$ measurements from PHI-FDT and HMI are known to be consistent \citep{2023A&A...673A..31S,2024A&A...685A..28M}. The noise level of the $B_{LOS}$ provided by the FDT of PHI is 7\,G \citep{2024A&A...685A..28M}.

The SO/PHI-FDT detector has a spatial sampling of 3''.75, corresponding to a linear scale of 761\,km on the solar surface, at Solar Orbiter’s closest perihelion of 0.28\,AU. This spatial scale varies with the spacecraft’s heliocentric distance, which ranged from 0.397\,AU to 0.93\,AU between 16 April and 18 July 2024. During this period, SO/PHI-FDT provided full-disk $B_{LOS}$ magnetograms at an almost regular cadence of 6 hours up to 4 July 2024 and 3 hours thereafter. 

The Full Sun Imager (FSI) of the Extreme Ultraviolet Imager \citep[EUI;][]{solo_eui} on board Solar Orbiter captures full-disk EUV images with a plate scale of 4''.4 per pixel. For our analysis, we use Level 2 data from Release 6 \citep{eui_data_release6}, specifically the 174\,\AA\ filtergram data.

Flare detections were provided by the Geostationary Operational Environmental Satellite (GOES) through the Space Weather Prediction Center (SWPC) of NOAA, for the Earth-facing side of the Sun, as well as by the Spectrometer/Telescope for Imaging X-rays \citep[STIX;][]{stix} onboard the Solar Orbiter. The latter detects solar X-ray emissions to study flare characteristics. The location information for flares observed by STIX was obtained from the  STIX science flarelist repository\footnote{\url{https://github.com/hayesla/stix_flarelist_science/}}, which also provides details of how the flarelist was generated. In this study, the GOES class of the flares observed only by STIX is estimated using the method described in \citet{stiefel_stix_flares}. For each flare, the peak flux measured by the STIX background detector is determined. Using the established correlation between this detector and GOES observations, the corresponding GOES class is estimated, associated with an 11\% uncertainty. As the STIX background detector is never covered by the attenuator, this approach allows for reliable classification of flares. This is important in the case of  large flares that trigger the attenuator \citep[e.g., > X1, see][]{stiefel_stix_flares}.

\begin{figure*}
  \centering
  \includegraphics[trim=0 330 0 0,  width=18cm]{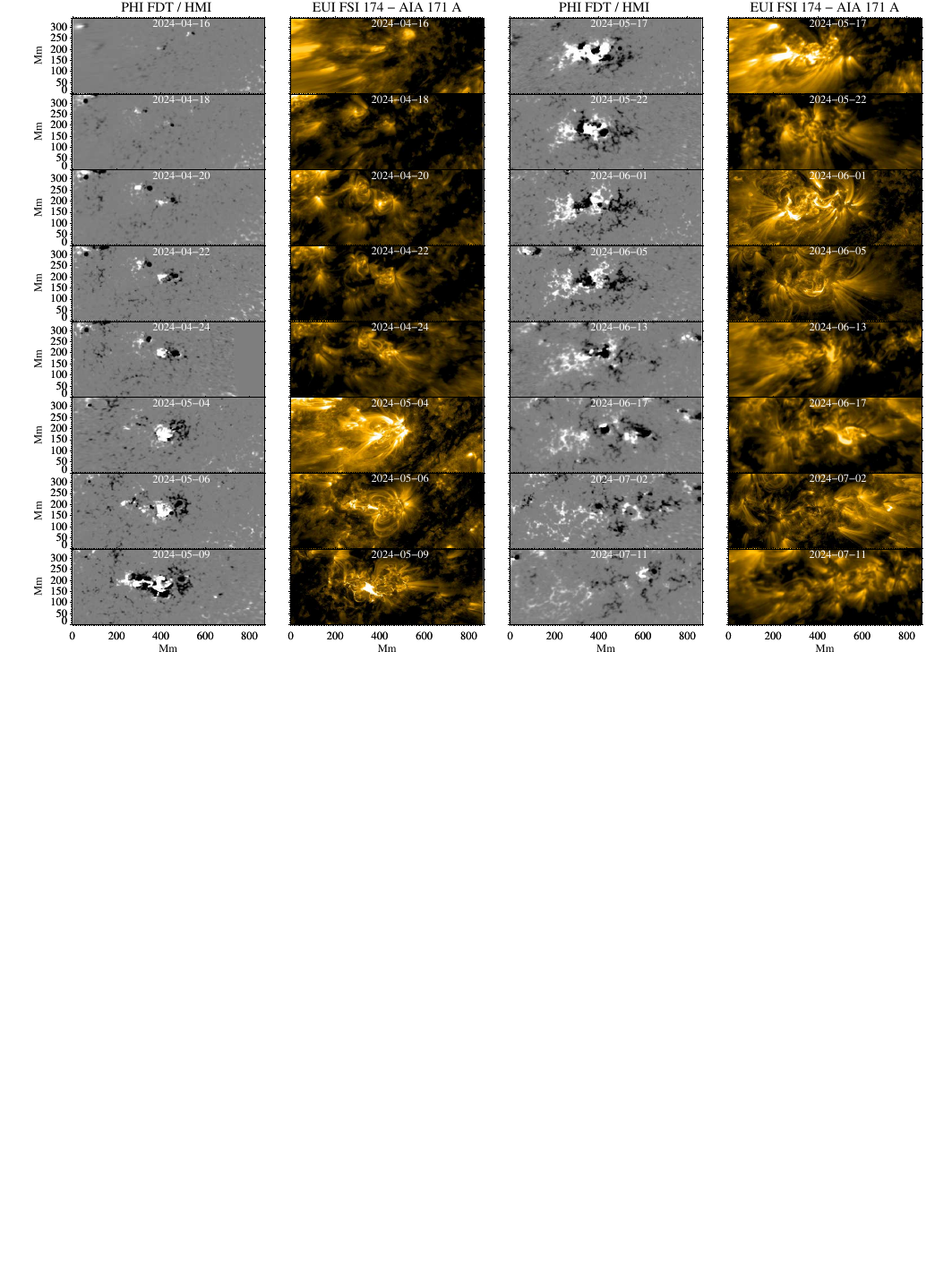}
  \caption{Snapshots of the evolution of NOAA\,13664/13697/13723 from 16 April 2024. Columns 1 and 3 show the HG-projected maps of $B_{LOS}$, while columns 2 and 4 show the corresponding maps of EUV emission at the 171-174\,\AA\ range. The magnetic field values have been scaled between $\pm$300\,G. An animated version of the magnetic field evolution is available online.}
       \label{fig:noaa_evo}
\end{figure*}

The HMI onboard the SDO provides photospheric line-of-sight and vector magnetograms, derived from spectropolarimetric observations in the Fe~{\sc i} 6173\,\AA\ line. For this study, we use the LOS magnetic field component, available at a cadence of 45\,s. The noise level of the $B_{LOS}$ provided by HMI is 10.2\,G \citep{2012SoPh..279..295L}. Our dataset is further supplemented by the Cylindrical Equal Area (CEA) version of the Space Weather HMI Active Region Patches \citep[SHARP;][]{bobra14}. This product includes the $B_r$, $B_p$, and $B_t$ components as well as $B_{LOS}$, all remapped to disk center coordinates. SHARP data were used as a reference to track the evolution of active region NOAA\,13664 and its recurrent appearances as NOAA\,13697 and NOAA\,13723.

The Atmospheric Imaging Assembly \citep[AIA;][]{aia} onboard SDO continuously images the upper solar atmosphere in UV and EUV wavelengths. We use data from the 171\,\AA\ channel, which provides images of the lower corona at a 12\,s cadence. These are compared with SO/EUI-FSI images obtained in the 174\,\AA\ channel. Although the two channels capture a similar coronal morphology, they image plasma with slightly different temperatures. The AIA\,171\,\AA\ channel is centered on the Fe~{\sc ix} line, which records emission from plasma at temperatures LogT (K)=5.8. The 174\,\AA\ channel of SO/EUI-FSI is centered on the Fe~{\sc x} line, and thus has a response to temperatures of around LogT (K)=6. Additionally, this channel has a larger emission contribution from plasma at transition region temperatures, in comparison to the 171\,\AA\ channel of AIA. 

To account for the varying spatial resolution of the SO/PHI-FDT $B_{LOS}$ maps over the Solar Orbiter's trajectory, all magnetograms were spatially binned to match the maximum effective plate scale, which corresponds to 7 HMI pixels. The same approach was applied to the SO/EUI-FSI filtergrams. As SO/PHI-FDT data are available at a maximum of 6-hour cadence, this cadence was imposed on all data sources. For each SO/PHI $B_{LOS}$ magnetogram, the nearest-in-time HMI $B_{LOS}$ map and SDO/AIA–SO/EUI-FSI filtergrams were selected, despite the higher native cadence of those instruments.

To approximate the radial magnetic flux from the LOS component, a $\mu$-angle correction was applied to each full-disk $B_{LOS}$ magnetogram. The SO/PHI-FDT and SDO/HMI magnetograms were deprojected into Heliographic Carrington coordinates. From these maps, cut-outs were extracted around the location of NOAA\,13664, ensuring that the selected boundaries also encompassed the extended regions corresponding to NOAA\,13697 and NOAA\,13723. The coverage of the solar disk in Heliographic Carrington longitudes provided by SDO and Solar Orbiter is illustrated in the right panel of Fig.~\ref{fig:orbits}. Accounting for the inevitable geometric foreshortening effects, the full active region can be seen in 90\% of the magnetograms.

\subsection{Magnetic field parameterization and analysis of the combined data set}
\label{parameters}
Several methods exist to quantify the non-potentiality (or complexity) of active regions and the non-potentiality of the photospheric magnetic field using its line-of-sight (LOS) component \citep{2021JSWSC..11...39G,2023AdSpR..71.2017K}. In this study, we focus on parameters associated with MPILs, which are key indicators of current-carrying magnetic fields and mark sites of flux emergence or cancellation. Specifically, we analyze the length of the main MPIL, the total length of MPILs within the field of view (FOV), and the associated magnetic flux located within 2\,Mm of the MPILs \citep{2018SoPh..293....9G}, as well as the R-value introduced by \citet{2007ApJ...655L.117S}. The latter quantifies the total unsigned magnetic flux within a 15\,Mm radius from the MPIL. These parameters are well-established predictors of imminent flaring and CME activity and correlate strongly with CME speed and acceleration \citep{kontogiannis19}.

As a reference for these calculations, we use parameters derived from the original HMI data, which are available only during the Earth-facing passage of active regions NOAA\,13664, 13697, and 13723. From the HMI $B_{LOS}$ maps, included in the SHARP cut-outs, we computed the MPIL-related parameters. These SHARP data products also include vector-field-derived quantities that characterize the magnetic field's non-potentiality. To provide context for the $B_{LOS}$-based parameters, we included three of these vector-derived parameters: the total unsigned magnetic flux, $\Phi$, and the total unsigned vertical electric current, $I_{z}$, representing fundamental physical measures of the region’s size and complexity, as well as the R-value computed from the $B_{r}$ component.

To extend our monitoring of magnetic non-potentiality from the Earth’s perspective and to connect with the previous analysis of NOAA\,13664 by \citet{kontogiannis2024}, we also calculated the non-neutralized electric current, $I_{NN,tot}$ \citep{geo_titov_mikic12}. This method segments the $B_{r}$ magnetogram into unipolar magnetic partitions, discards the smallest ones, and computes the electric current for the remaining partitions. Electric currents that significantly exceed the values expected from error propagation and numerical effects are classified as non-neutralized \citep[for methodological details, see][]{kontogiannis17,2024ApJ...970..162K}. 

The same magnetic partitioning method used to calculate $I_{NN,tot}$ was also applied on the combined, HG-projected $B_{LOS}$ maps to track the evolution of the total magnetic flux and the magnetic area of each region.


\begin{figure}
   \centering
   \includegraphics[trim=0 30 0 0,  width=9cm]{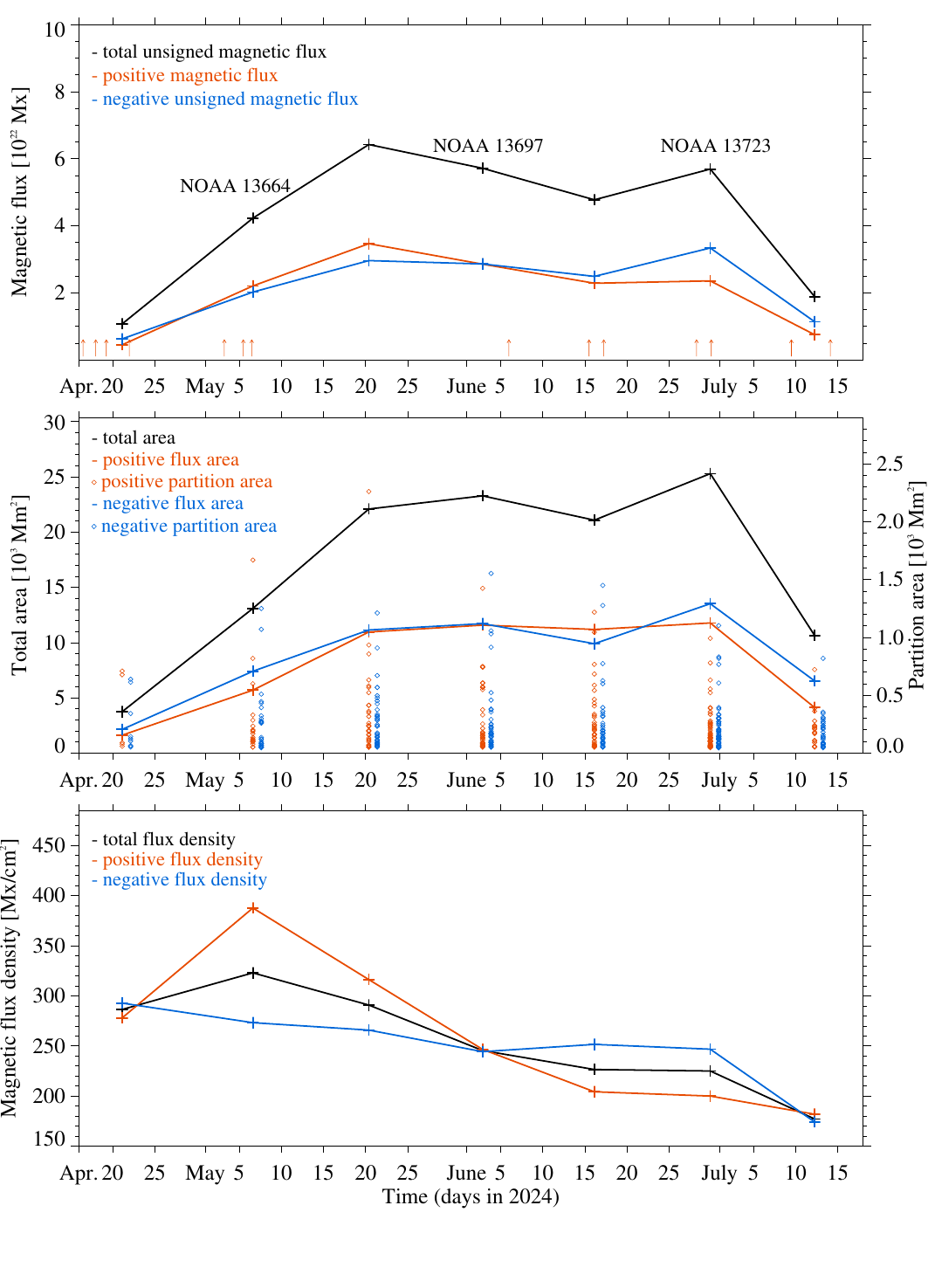}
      \caption{From top to bottom: summary evolution of the magnetic flux, magnetic area and magnetic flux density (magnetic flux over magnetic area) of the region, when it was crossing the central meridian. Black curves represent the total unsigned magnetic flux (and corresponding area), while blue and red represent positive and negative flux-related quantities, correspondingly. The blue and red diamonds in the middle panel represent the individual areas of the positive and negative magnetic partitions. Red arrows in the top panel indicate the times of flux emergence episodes in the FOV.}
        \label{fig:flux}
\end{figure}

\begin{figure*}
   \centering
   \includegraphics[width=18cm]{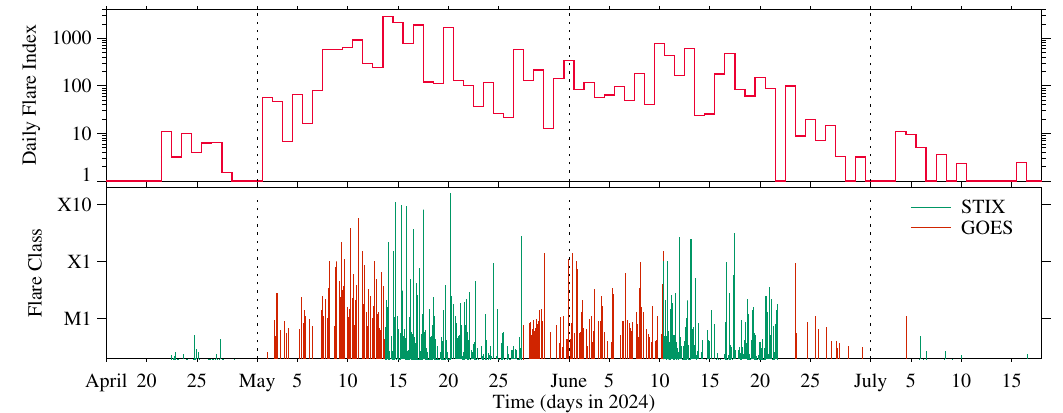}
      \caption{The daily flare index (top) and the individual flares (bottom) of of the NOAA\,13664/13697/13723 activity complex, as detected by GOES (red) and STIX (green). Overplotted are also gray arrows that indicate the conspicuous flux emergence episodes (see text.)}
        \label{fig:flare_index}
\end{figure*}

\section{Results}
\label{results}
\subsection{Morphological evolution}
\label{magnetic_evo}

When active region NOAA\,13664 first appeared on the Earth-facing side of the Sun, it was already well-developed. It featured a prominent leading sunspot of negative polarity and an extended trailing region of positive polarity, containing numerous smaller sunspots and at least one $\delta$-spot region \citep{kontogiannis2024}. The region retained the $\delta$-type for two more rotations, as NOAA\,13697 and NOAA\,13723, although the latter decayed into $\beta$-type soon after its central meridian passing. With the combined observations from the SO/PHI-FDT and SDO/HMI, we can now trace the origins of this activity complex, which was located on the far side of the Sun, and monitor its evolution almost continuously for approximately three months, spanning more than three full solar rotations. A summary of the region’s magnetic field and coronal emission evolution from emergence to decay is presented in Fig.~\ref{fig:noaa_evo}.

The region originated as a small bipole, first observed on 16 April 2024 at (x,y) $\sim$ (400,250), emerging within a predominantly negative-polarity area (Fig.~\ref{fig:noaa_evo}, column 1, row 1). An ephemeral region was already present to the northwest, and both bipoles were associated with coronal EUV emission from small-scale loops. While that ephemeral region gradually decayed, another emerged to the northeast by 17 April. In the following days, both bipoles grew and exhibited increased EUV emission, indicating the formation of magnetic connections between the opposite polarities. At least three additional flux emergence events occurred within the main region, further expanding it and increasing its EUV output (Fig.~\ref{fig:noaa_evo}, columns 1-2, rows 3-5). These early emergence events likely triggered the weak flaring activity, with multiple B- and C-class flares, recorded between 22 and 28 April 2024 by STIX (see also Fig.~\ref{fig:flare_index}).

Between 26 and 29 April 2024, the region was in neither instrument's FOV (Fig.~\ref{fig:orbits} right panel, dotted lines). It then appeared for the first time in the HMI FOV, and was later labeled NOAA\,13664 (Fig.~\ref{fig:noaa_evo}, column 1, row 6). It had further developed, showing pronounced complexity and stood out in EUV, showing many bright, tangled coronal loops. New magnetic flux continued to emerge along the main MPIL, and parasitic negative polarities exhibited shearing motions around the trailing positive polarity (see online material). These features contributed to increased magnetic complexity at the trailing part, coinciding with enhanced flaring during the first week of May 2024, including several C- and M-class flares. By this time, the region's configuration was sufficiently complex to support major flaring, as also shown in \citet{kontogiannis2024}. Three major flux emergence episodes followed, starting on 5 May, which resulted in the merging of the three flux systems and the development of a highly complex magnetic structure at the trailing part, seen in the bottom left panel of Fig.~\ref{fig:noaa_evo}. This new configuration featured multiple strong, alternating polarities and persistent shearing flows, driving intense flaring activity. A C8.8 and X2.3 class flares, on 6 and 9 May 2024, respectively, can be seen in the last two panels of the second column in Fig.~\ref{fig:noaa_evo}. Multiple X-class flares occurred during the period between 08 and 12 May 2024, including some of the strongest of Solar Cycle 25, causing the geomagnetic storm of that period \citep[see also][]{2025ApJ...988..108D}.

As the region re-entered the SO/PHI-FDT FOV after 14 May, the leading negative polarity remained discernible, but the rest of the region 
exhibited a ``braided'' morphology, with several compact positive polarities forming a W-shaped pattern surrounded by negative polarities (Fig.~\ref{fig:noaa_evo}, 1st row, 3rd column). The middle part of the region exhibited strong, large-scale rotating and shearing motions and intense flux cancellation, which led to the X16.5 flare, on 20 May 2024, the strongest flare produced by the region. After 25 May the gradual decay of the region started, as now the magnetic polarities had become more diffuse and spatially extended. 

When the region rotated once again into the Earth view, it was designated NOAA\,13697 and it consisted primarily of four discrete magnetic zones arranged in a positive-negative-positive-negative pattern (Fig.~\ref{fig:noaa_evo}, 3rd row, 3rd column). Although some polarity mixing and interactions persisted, the region was now clearly less compact. Flaring activity continued, but only marginally reached X-class levels. Signs of magnetic decay became more evident through flux cancellation and fragmentation at the outer boundaries. The trailing positive and leading negative polarities began to break apart into smaller magnetic concentrations, while the central negative polarity grew more dominant. In the corona the main part of the region became less and less pronounced in comparison to the rest of the FOV

Upon its return into the SO/PHI-FDT FOV, the region appeared as a disorganized set of dispersed polarities (Fig.~\ref{fig:noaa_evo}, 5th-6th row, 3rd column), although magnetic interactions remained strong. A brief resurgence in flaring occurred in mid-June, likely driven by additional episodes of flux emergence. By the time the region re-entered the Earth-facing side as NOAA\,13723 (Fig.~\ref{fig:noaa_evo}, 7th row, 3rd column), it resembled an ensemble of smaller polarities, only loosely echoing its earlier structure. Some flux emergence was still observed, mainly at the periphery, particularly ahead of the region (the leading part of the region was now brighter than the following), but the overall complexity and flaring activity had markedly decreased. Following its final rotation into the SO/PHI-FDT FOV on 5 July 2024, the region was characterized by dispersed, small-scale polarities (see also Fig.~\ref{fig:flux} and Sec.~\ref{sec:magflux_evo}). Flaring activity had diminished to only sporadic B- and C-class events, mainly at the periphery of the region, marking the end of the region’s significant contribution to solar activity during this observational period. 

After 18 July 2024, this solar region rotated out of the SO/PHI FOV. When it reappeared in HMI, the region was now occupied by scattered, small scale positive and negative magnetic fields. The magnetic activity had shifted towards the right edge of the FOV, where another region, NOAA\,13768 was emerging. 

\subsection{Evolution of the magnetic flux}
\label{sec:magflux_evo}
The overall evolution of the active region can be characterized in terms of its total unsigned magnetic flux, area, and magnetic flux density (Fig.~\ref{fig:flux}). These quantities were measured only for the significant magnetic partitions present at the time of central-meridian crossing, rather than across the full time series, in order to minimize the impact of projection and foreshortening effects on the corresponding trends. The complete time series of the total unsigned magnetic flux will be discussed in Sec.~\ref{complexity}.

In terms of central meridian crossings, the region reached its peak total unsigned magnetic flux on 20 May, more than a month after the initial flux emergence (top panel of Fig.~\ref{fig:flux}). A gradual decline followed, interrupted by a secondary peak at the beginning of July. The decay was steep during the last two weeks of the observing window. The individual contributions from the positive and negative polarities exhibited similar trends, with the flux imbalance reversing sign twice during the region’s lifetime, first at the end of April, in favor of the positive polarity during the peak phase, and again at the beginning of June, when the negative polarity became dominant. The timing of 16 individual flux emergence episodes, which could be detected visually, is also indicated in the top panel of Fig.~\ref{fig:flux}. Initially, these events were concentrated in the core of the region and contributed to its growth. In later phases, emergence occurred primarily in the periphery and caused less substantial increases in the total flux within the FOV. 

The total area occupied by magnetic flux (middle panel of Fig.~\ref{fig:flux}) followed a broadly similar trend: a steady increase until mid-May, a one-month long plateau, followed by a secondary rise until late June, and then a rapid decline during the last two weeks. The areas associated with the positive and negative polarities evolved in parallel. Early in the evolution, the negative polarity covered a larger area, as the positive polarity was more compact. Around the midpoint of the region's lifetime, the two polarities were approximately balanced in spatial extent. Toward the end, however, the negative polarity again became more extended, consistent with the observed flux imbalance.

The size distribution of individual magnetic partitions (middle panel of Fig.~\ref{fig:flux}, red and blue diamonds) also varied throughout the evolution of the region. Initially, the region comprised a few small partitions (14), but their number and size steadily increased over time. Their maximum size peaked during the most active phase (around 20 May) while their number reached a local maximum (90) during the central meridian crossing of NOAA\,13697. Then the maximum size of the partitions progressively decreased, although their number reached its maximum value (123), during the central meridian crossing of NOAA\,13723 (see also Fig.~\ref{fig:noaa_evo}, 7th and 8th row, 3rd column). Eventually, many partitions fell below the detection threshold, contributing to the rapid apparent decline in magnetic area. By the end of the observational period, the field of view was dominated by mixed, small-scale magnetic elements (57, during the final central meridian crossing).

In contrast, the magnetic flux density, defined as the ratio of total magnetic flux to its associated area, displayed a distinct evolutionary pattern. It peaked in early May and subsequently decreased. While the positive polarity followed this general trend, its decline was steeper, whereas the decline of the negative polarity was consistent from the beginning, but less steep. Notably, the negative polarity was the first to form a stable sunspot, but eventually persisted into the late stages of the region’s evolution, potentially explaining its differing behavior.

\subsection{Flaring activity}
\label{flaring}

We compiled a comprehensive flare catalog using GOES flare events attributed to NOAA 13664, 13697, and 13723, supplemented by STIX detections that corresponded to the far-side manifestations of these active regions. Over the 94-day period, STIX and GOES recorded a total of 969 flares, comprising 38 X-class, 146 M-class, 527 C-class, and 258 B-class events. The total flare output is illustrated in Fig.~\ref{fig:flare_index}, both as individual events and as a time series of the daily flare index. The latter was computed following the method of \citet{2005ApJ...629.1141A}, where flare classes were first converted into peak soft X-ray flux values, and then integrated over consecutive one-day bins, beginning on 16 April 2024.

As also discussed in Sec.~\ref{magnetic_evo}, during the first stages of emergence the region was associated with weak flaring. The first recorded event, a B4-class flare, occurred on 22 April 2024, four days after the start of observations. This was followed by a series of B- and C-class flares, with magnitudes reaching up to C4.1 on 23 April. By 2 May, the region was already producing M-class flares with increasing frequency. The first X-class flare occurred on 8 May 2024, marking the onset of a highly active phase characterized by frequent X- and M-class events. Peak activity was reached around 14 May in terms of daily flare index, and on 20 May in terms of individual flare magnitude, when the strongest event, an X16.5-class flare, was recorded by STIX. Although the intensity and frequency of flares declined thereafter, activity remained elevated for nearly four additional weeks. The final X-class flare was observed on 17 June 2024, after which the flare output dropped significantly.

Assuming the X16.5 flare on 20 May represents the main eruptive event, it coincided with the region’s peak total unsigned magnetic flux (as measured during central meridian crossings, Fig.~\ref{fig:flux}). The two-week integrated flaring output of NOAA\,13664 was only up to 33\% of the corresponding activity during the two weeks that followed (recorded only by STIX), whereas the activity of NOAA\,13697 and 13723 reached up to 17\% and 1.5\% of that, correspondingly. Overall, 63\% of X-class, 50\% of M-class, 38\% of C-class, and 53\% of B-class flares occurred prior to the main event. Thus, over half of the flare events, including the most powerful ones, took place during approximately the first one-third of the active region’s lifetime, even though the region continued to be flare-productive well beyond that point. However, in comparison to other studied activity complexes, which exhibited notable flaring only during one of their rotations \citep{1999ESASP.446..663V,2021RAA....21..312M}, NOAA\,13664/13697/13723 was consistently active up until its decay.

\begin{figure*}
   \centering
   \includegraphics[width=18cm]{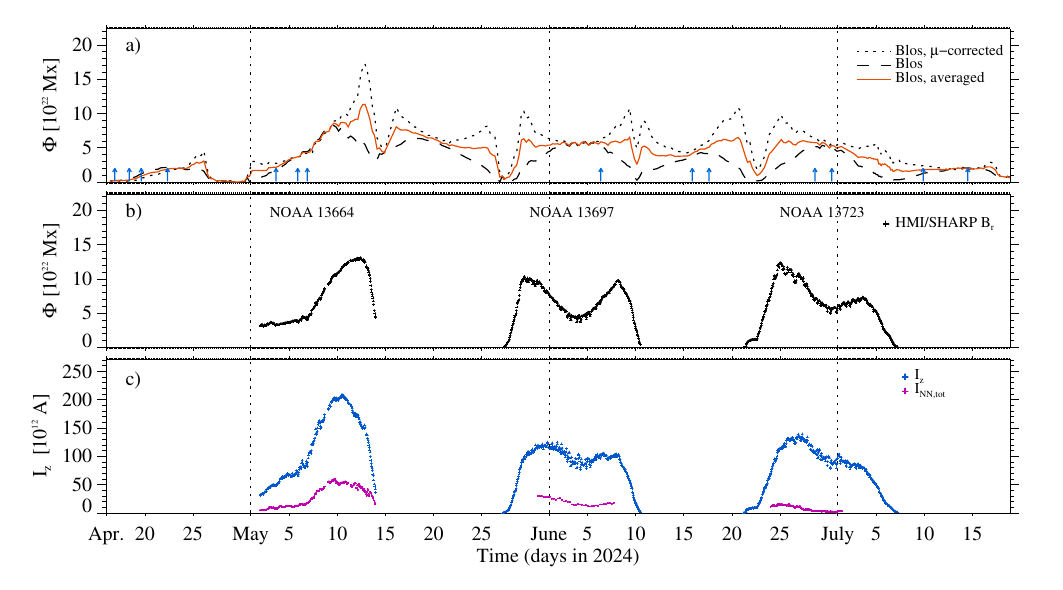}
      \caption{a) The total unsigned magnetic flux calculated from the collated data set, for both $\mu$-angle corrected and uncorrected LOS magnetograms (dotted and dashed lines respectively). The red line represents the average between the two curves. The blue arrows indicate the onset times of flux emergence events, also shown in Fig.~\ref{fig:flux}. b) The total unsigned magnetic flux provided from the SHARP and calculated using the radial component, $B_{r}$, of the magnetic field. c) the total unsigned vertical electric current provided from the SHARP (blue crosses) and the total unsigned non-neutralized electric current calculated for the same data (magenta crosses). Dashed, vertical lines mark the first day of each month, to facilitate comparison. }
        \label{fig:flux_current}
\end{figure*}

\subsection{Evolution of magnetic non-potentiality}
\label{complexity}
\begin{figure*}
   \centering
   \includegraphics[trim=0 15 0 0, width=18cm]{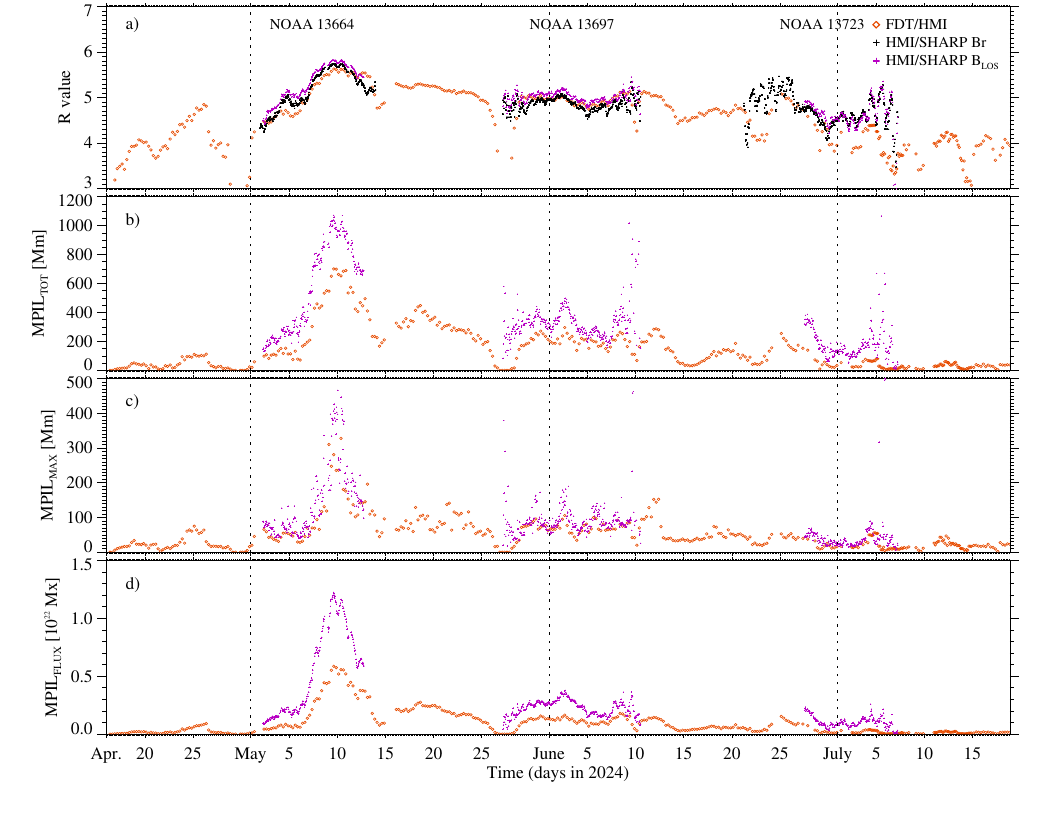}
      \caption{Same as Fig.~\ref{fig:flux_current} for a) the R-value calculated from the SHARP $B_{r}$ and $B_{LOS}$ components (black and purple crosses) and from the collated data set (red diamonds). b-d) the total length of MPILs, the length of the main MPIL and the magnetic flux associated with MPILs, calculated from the SHARP $B_{LOS}$ and the collated SDO/HMI -- SO/PHI-FDT data set (red diamonds and purple crosses respectively). Dashed lines mark the first day of each month, to facilitate comparison. }
        \label{fig:mag_params}
\end{figure*}

The combined datasets from SDO/HMI and SO/PHI-FDT enable an almost continuous characterization of the non-potentiality, or magnetic complexity, of the region under study. All relevant parameter time series (see Sec.~\ref{parameters}) are presented in Fig.~\ref{fig:flux_current} and~\ref{fig:mag_params}. It is important to note that the goal of the following discussion is not to perform a detailed statistical analysis of the parameter values, which would require a large sample of active regions with varying sizes and activity levels, but rather to demonstrate the potential of using such combined datasets in both exploratory and operational contexts \citep[see also][]{2021ApJS..256...26B,2024ApJ...972..169K}.

A general observation concerning the temporal evolution of magnetic non-potentiality is that it closely follows the trend of the flare index and flaring activity shown in Fig.~\ref{fig:flare_index}. Each parameter captures different aspects of non-potentiality and is affected to varying degrees by projection effects, foreshortening, and instrumental biases  \citep{kontogiannis18}. Despite being derived from a combined dataset, the parameters exhibit a consistent behavior over the 94-day continuous monitoring period. We now examine these parameters in greater detail, bearing in mind that some are derived from HMI vector magnetograms (which are provided for context), while others rely solely on the line-of-sight component of the magnetic field.

In Fig.~\ref{fig:flux_current}a we present the time series of the total unsigned magnetic flux for the complete duration of the combined data set. If the full-disk $B_{LOS}$ magnetograms are not corrected for the projection angle (Fig.~\ref{fig:flux_current}a, dashed line), then the total unsigned magnetic flux is decreasing towards the limb \citep[see also][]{2021ApJS..256...26B}. Using the $\mu$-angle correction alleviates the problem for small viewing angles \citep{2017SoPh..292...36L} but leads to an increase of the total unsigned magnetic flux towards the limb (Fig.~\ref{fig:flux_current}a, dotted line). This inherent difficulty to objectively determine the magnetic flux on any position on the solar disk was the reason why we constrained our analysis so far on the central meridian crossings of the active region (Fig.~\ref{fig:flux}). However, on top of the periodic variation imposed by varying position on the solar disk we can still see an overall trend which is in agreement with the behavior described in Sec.~\ref{magnetic_evo} and Fig.~\ref{fig:flux}: an initial increase of the total magnetic flux during the first month, peaking in mid-May, followed by a gradual decline, which accelerates in the final two weeks. In the same panel we also include another estimate of the total unsigned magnetic flux (Fig.~\ref{fig:flux_current}a, red line), which is calculated from the average maps of the original and the $\mu$-angle corrected $B_{LOS}$. This version seems to represent the temporal evolution of the magnetic field of the region, with less pronounced projection effects. At the end of each two-week observational window from either instrument, the magnetic flux and other parameters drop abruptly as the region becomes partially obscured near the limb. Notably, both time series show clear signatures of flux accumulation following flux emergence episodes (Fig.~\ref{fig:flux_current}a), blue arrows). This is evident after 17 April, in early May 2024, when the region underwent rapid development and approached its peak flux, and again after 14 June, when renewed emergence led to a secondary peak in early July. Later emergence events in July had minimal impact on the total flux. 

The total unsigned magnetic flux provided by the SHARP data (Fig.~\ref{fig:flux_current}b) also exhibits a wavy pattern, which is due to the spatially dependent sensitivity of HMI \citep{2014SoPh..289.3483H}. Overall, considering the different systematic effects in the two datasets, the different vantage points and varying distance of Solar Orbiter, we conclude that the combined $\Phi$ values, provide a coherent summary of the active region's magnetic evolution.  

The vertical electric current (Fig.~\ref{fig:flux_current}c, blue crosses) is less sensitive to systematic effects of HMI, as it is calculated from spatial derivatives of the horizontal magnetic field components and emphasizes regions of strong, sheared fields \citep{ravindra11,geo_titov_mikic12,torok14,dalmasse15,2022SoPh..297...59K}. This emphasis is even more pronounced in the $I_{NN,tot}$ values (Fig.~\ref{fig:flux_current}c, magenta crosses), which specifically stem from the current-carrying portion of the region \citep{2024ApJ...970..162K}, accounting for the generally lower values of this parameter in comparison to $I_{z}$. As previously noted by \citet{kontogiannis2024}, both $I_z$ and $I_{\mathrm{NN,tot}}$ were already elevated when the region first appeared on the Earth-facing solar disk, indicative of significant shear and ongoing flux emergence. A sharp increase in both quantities after 7 May corresponded to new flux emergence and the subsequent strong interaction between magnetic flux systems \citep{Jarolim2024,wangrui2024}. Elevated current levels persisted during the following two solar rotations, albeit at lower levels consistent with a decaying region. The flux emergence events on 14 and 16 June likely contributed to an increase in vertical current, explaining why NOAA\,13723 rotated into view with a higher total current than NOAA\,13697, indicating the presence of still significant shear. The electric current began a steady decline after 1 July. Notably, the $I_{\mathrm{NN,tot}}$ time series illustrate more clearly the decreasing magnetic complexity over time and across the three rotations, with values decreasing, from NOAA\,13697 to NOAA\,13723. This is attributed to the fact that the non-neutralized electric current is more closely tied to regions of strong shear and is less influenced by the spatial extent of the active region and dispersed magnetic polarities, which, though they may contribute to the total current, they tend to be largely neutralized.

The MPIL-related quantities, derived from the combined dataset and presented in Fig.~\ref{fig:mag_params}, offer a coherent and consistent characterization of the magnetic field's non-potentiality throughout the 94-day observation period. These parameters align well with the trends observed in total magnetic flux and vertical electric current, while also providing complementary information, particularly valuable during intervals when the region was located on the far side of the Sun. 

The $R$-value, which quantifies the magnetic flux within 15\,Mm of MPILs, increased markedly during the initial flux emergence episodes (Fig.~\ref{fig:mag_params}a, red diamonds). This increase is likely exaggerated near the Solar Orbiter-visible limb due to projection effects, and then declines during the observational gap around 26--29 April. Upon appearance within the SDO/HMI FOV, the R-value stabilized around 4.5 and began to rise, signaling the onset of significant flaring activity \citep{2007ApJ...655L.117S}. During the dynamic evolution that followed, R-value peaked at 5.8, exceeding the values recorded for all other active regions so far \citep[1.5 -- 5.5;][]{2018SoPh..293....9G,Jaswal2025}. Then it declined gradually, yet remained elevated through the end of May, coinciding with the period of most intense flaring. The R-value continued to vary intermittently in response to additional flux emergence episodes, indicating localized increases in magnetic complexity that punctuated the region’s otherwise steady decay.

The remaining three MPIL-related parameters exhibit broadly similar temporal evolution to the R-value, though with notable distinctions arising from their differing definitions and the fact that R-value is expressed in a logarithmic scale. The parameter $MPIL_{TOT}$ (Fig.~\ref{fig:mag_params}b, red diamonds), representing the total length of all individual MPILs within the region, displays prominent peaks following flux emergence events and, most notably, during episodes of strong magnetic interactions. This accounts for the marked increase after 5 May 2024, as well as the consistently elevated $MPIL_{TOT}$ leading up to the X16.5 flare on 20 May. A gradual declining trend ensued, punctuated by several conspicuous peaks, until a sudden drop is observed after 5 July 2024.

The $MPIL_{MAX}$ parameter (Fig.~\ref{fig:mag_params}c), which tracks the length of the longest MPIL at each time step, follows a similar trend but exhibits increased variability and sharper peaks. This behavior is expected, as the parameter reflects the most dominant MPIL at any given time. In a highly dynamic and fragmented active region such as NOAA\,13664–13697–13723, where multiple MPILs of comparable strength evolve concurrently, the identity of the ``main'' MPIL can shift depending on the details of ongoing interactions.

Finally, the magnetic flux associated with MPILs, $MPIL_{FLUX}$ (Fig.~\ref{fig:mag_params}d, red diamonds), differs from the R-value in that it measures flux within a much narrower area around the MPILs. Its evolution largely mirrors that of the other two parameters, though the time series is comparatively smoother. This is because changes in MPIL length between successive LOS magnetograms do not always coincide with significant variations in the magnetic field strength in their immediate vicinity.   

In Fig.~\ref{fig:mag_params}, we also included values of the examined parameters calculated from the deprojected $B_{LOS}$ maps provided with SHARP, as well as R-values derived from the radial magnetic field component $B_{r}$ (represented by purple and black crosses, respectively). These SHARP $B_{LOS}$ maps differ from the deprojected SO/PHI-FDT – SDO/HMI dataset in two key aspects: (1) their pixel size is approximately seven times smaller, and (2) they are deprojected using a Cylindrical Equal-Area (CEA) projection, rather than a heliographic (HG) Carrington projection. As a result, the magnetic flux values in the combined SO/PHI-FDT - SDO/HMI dataset are generally lower and smoother, owing to the larger pixel size. Additionally, in the HG projection, the area of each pixel varies with latitude, whereas in the CEA projection it remains constant. However, this difference is of limited relevance here, as we are not comparing active regions found at different latitudes.

The comparison between the SHARP and combined datasets reveals the following: (1) all parameters show consistent overall trends, particularly when the region is located away from the limb; (2) the R-values differ only marginally under these conditions; (3) the length of the main polarity inversion line, $MPIL_{MAX}$, shows the closest agreement between the datasets, likely because it reflects only the longest MPIL in the field of view, thus minimizing cumulative differences; and (4) $MPIL_{FLUX}$ exhibits the greatest discrepancies, probably due to its dependence not only on MPIL geometry but also on local magnetic flux values, which are more sensitive to resolution and deprojection effects. 

The combined SDO/HMI and SO/PHI data set facilitates the study of the long-term evolution of the non-potentiality of active regions, not only for exploratory, but also for operational flare and CME-prediction purposes. All $B_{LOS}$-derived parameters of Fig.~\ref{fig:mag_params} are positively correlated with the flare index, with the calculated Spearman (rank) coefficients indicating a significant association via a non-linear monotonic relation (Table~\ref{tab:correl_coeffs}). The strongest correlation is exhibited by the $R$-value (Spearman correlation coefficient equal to 0.76), followed by $MPIL_{FLUX}$ (0.74), the total unsigned magnetic flux (0.74), $MPIL_{TOT}$ (0.72) and $MPIL_{MAX}$ (0.68).

\begin{table}[htbp]
    \centering
    \caption{Spearman correlation coefficients between flare index and the magnetic field parameters presented in Fig.~\ref{fig:flare_index} and \ref{fig:mag_params} respectively. The coefficients are calculated for the 94 days of coverage.}
    \label{tab:correl_coeffs}
    \begin{tabular}{lcc}
        \hline
        \hline
        Parameter & Pearson  &  Spearman (rank)  \\
        \hline
        $\Phi$            & 0.44 &  0.74   \\
        $R$-value       & 0.33 &  0.76   \\
        $MPIL_{TOT}$     & 0.47 & 0.72  \\ 
        $MPIL_{MAX}$     & 0.29 & 0.68  \\
        $MPIL_{FLUX}$     & 0.45 & 0.74  \\
        \hline
    \end{tabular}
\end{table}


\section{Conclusions and Discussion}

We presented a near-continuous, long-term study of the super active region NOAA\,13664-13697-13723, combining multi-wavelength observations from two vantage points. Over a period of 94 days, we tracked all evolutionary stages of the region, from its emergence on the far side of the Sun to its eventual decay, including its flaring activity and the evolution of its non-potentiality.

To our knowledge, this is the first time that all evolutionary stages of NOAA\,13664 have been continuously monitored in such detail. This undertaking was previously impossible for extended and complex regions like this, but it is now feasible due to the regular, favorable geometric alignments of Solar Orbiter and SDO \citep[see also][]{Finley2025}. In our case, the two spacecraft were positioned almost in antithesis, enabling nearly uninterrupted monitoring of the region.

The active region formed through repeated episodes of magnetic flux emergence. Over approximately twenty days, a large, primarily bipolar configuration developed, including some parasitic polarities. In the following two weeks, intense flux emergence occurred nearby, and subsequent merging with the pre-existing structure led to an extremely complex configuration. Several sub-regions exhibited strong magnetic shear, eventually forming a compact, “braided” structure.

At its peak, large-scale rotating and shearing motions of the negative polarities disrupted this compact structure. The region then entered a prolonged decay phase lasting at least one month. During this time, both the total magnetic flux and the level of non-potentiality steadily decreased, although the region retained significant residual complexity and showed intermittent flux emergence. Following a final, brief resurgence in magnetic flux, the region underwent a rapid decay during the last two weeks of its observed lifetime. It is likely that the persistent complexity of the magnetic field led to a faster decay, through dissipation events.

The flaring activity of the region closely followed the evolution of its magnetic flux and non-potentiality. It peaked around the time of maximum size and complexity, then gradually declined, with brief resurgences likely responding to episodes of flux emergence. Notably, although the overall flare activity, measured in terms of integrated output, peaked on 14 May 2024, the strongest flare occurred six days later, on 20 May, following two days of relatively lower activity. Overall, the region exhibited remarkably persistent and intense flaring that lasted for almost two months and was notable for longer than two solar rotations, producing 969 events. A comparison to the few other regions that have been studied over long term (but with the inevitable gaps due to the solar rotation) shows that this persistent flaring activity of NOAA\,13664-13697-13723 is unique, but further studies are needed to support a statistically significant conclusion. Future works on combined observations from many vantage points can provide more details on the long-term evolution of active regions and complexes and their flaring activity \citep{Finley2025}. Our results support, however, that the observed behavior is due to the significant magnetic non-potentiality sustained by the region, indicating the presence of substantial free magnetic energy over more than two months.  

Strong magnetic interactions observed in the photospheric magnetic field are often interpreted as surface manifestations of highly complex sub-photospheric structures, with segments emerging in close spatial and temporal proximity \citep{2013ApJ...764L...3C,2017ApJ...850...39T,2024ApJ...973L..31R,2025ApJ...988..108D}. The repeated emergence episodes observed in NOAA\,13664, primarily during its emergence phase, and the development of an extended, highly complex, and persistent configuration, strongly suggest the presence of a highly twisted and deformed sub-photospheric flux system. A notable characteristic of this region is that some of the most defining emergence episodes, particularly those in May 2024, occurred nearly two weeks after the initial emergence. This supports the hypothesis that different branches or segments of the same sub-photospheric flux tube, each with varying degrees of twist, emerged at different speeds and times before merging at the photosphere. Comprehensive multi-spacecraft, long-duration observational studies, combined with advanced modeling, will be essential to further elucidate the origins and evolution of such complex active regions in the future.

This study also highlights the future potential of combining magnetogram data from spacecraft positioned at different vantage points for both active region evolution studies and the forecasting of their eruptive activity. Due to the Solar Orbiter’s variable distance from the Sun and its lower spatial resolution, particularly outside perihelion, a downgrading of the HMI $B_{LOS}$ maps and a deprojection on a common coordinate system was required. Despite this processing, which certainly smooths out features, the derived complexity parameters showed remarkable consistency. The nearly continuous time series of magnetic parameters produced by the combined datasets of SO/PHI-FDT and SDO/HMI correlated strongly with the region's flaring activity. Perhaps the use of resolution enhancement techniques \citep{2024SoPh..299...36X} on SO/PHI-FDT in the future could lead to smaller differences between data sets and even more uniform image series and parameter sets.

A logical next step would be to derive non-potentiality parameters such as the one presented in Fig.~\ref{fig:mag_params}, both from $B_{LOS}$ and the full magnetic field vector, using joint datasets from the two spacecraft. This could lead to more extended datasets, to which the efficiency of those parameters as flare and CME predictors could be tested. This is particularly timely, given the ongoing monitoring of the Sun during strong solar activity. At the same time, the data gaps that persisted in our observations underscore the importance of observations away from the Sun-Earth line, such as with future missions like Vigil\footnote{\url{https://www.esa.int/Space_Safety/Vigil}}, and, most importantly multi-vantage point observations, as envisioned in proposed future missions \citep{2024JASTP.25406165G,2023BAAS...55c.333R}. In such a scenario, solar magnetic monitoring would become truly continuous, with projection effects significantly minimized.



\begin{acknowledgements}
    We would like to thank the anonymous referee who provided comments that improved the content and clarity of the manuscript. Solar Orbiter is a space mission of international collaboration between ESA and NASA, operated by ESA. We are grateful to the ESA SOC and MOC teams for their support. The German contribution to SO/PHI is funded by the BMWi through DLR and by MPG central funds. The Spanish contribution is funded by AEI/MCIN/10.13039/501100011033/ and European Union “NextGenerationEU”/PRTR” (RTI2018-096886-C5,  PID2021-125325OB-C5,  PCI2022-135009-2, PCI2022-135029-2) and ERDF “A way of making Europe”; “Center of Excellence Severo Ochoa'' awards to IAA-CSIC (SEV-2017-0709, CEX2021-001131-S); and a Ramón y Cajal fellowship awarded to David Orozco-Suarez. The French contribution is funded by CNES. The STIX instrument is an international collaboration between Switzerland, Poland, France, Czech Republic,
Germany, Austria, Ireland, and Italy. Data from SDO/HMI are courtesy of NASA/SDO and the AIA, EVE, and HMI science teams and are publicly available through the Joint Science Operations Center at the \url{jsoc.stanford.edu}.)
\end{acknowledgements}

%
%

\bibliography{references}{}
\bibliographystyle{aa}

\begin{appendix}
\section{Emergence events}

\begin{figure*}[b]
   \centering
   \includegraphics[width=18cm]{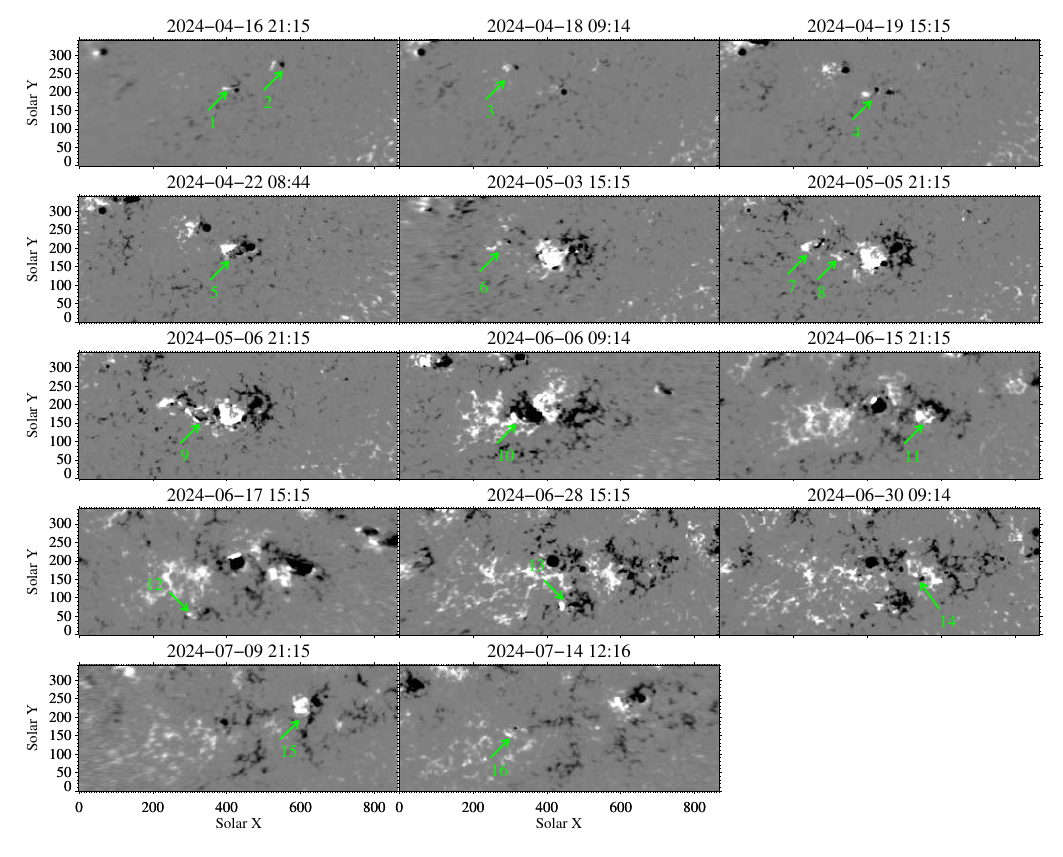}
      \caption{Emergence Events. }
        \label{fig:emergence_events}
\end{figure*}

In Fig.~\ref{fig:emergence_events} we present snapshots with the emergence events that could be detected visually. Out of these events, No.1 and 2. were present at the beginning of the observations, but No.2 was already decaying. Overall, the strongest events were seen during the first one-third of the regions lifetime (No.5-9), and they were the ones that shaped it drastically \citep[see also][]{kontogiannis2024}. Strong events were also detected during the first month of the decaying phase (No.10-11).

\end{appendix}
\end{document}